\documentclass[twocolumn,prb,showpacs,multicol,amsmath,amssymb]{revtex4}
\usepackage[dvips]{graphicx}
\usepackage{graphicx}
\usepackage{dcolumn}
\usepackage{bm}
\usepackage{graphics}
\usepackage{epsfig,color}

\newcommand{\be}{\begin{equation}}
\newcommand{\ee}{\end{equation}}
\newcommand{\bea}{\begin{eqnarray}}
\newcommand{\eea}{\end{eqnarray}}

\newcommand{\bS}{\bf S}

\begin{document}

\title{Metamagnetic phase transition in the Ising plus Dzyaloshinskii-Moriya model}
\author{M. R. Soltani$^{1}$, S. Mahdavifar$^{2}$, Alireza Akbari$^{3,4}$, A. A. Masoudi$^{1,5}$}
\affiliation{$^{1}$Physics Research Center, Science Research Branch, Islamic Azad University, 19585-466, Tehran, Iran \\
$^{2}$ Department of Physics, University of Guilan,41335-1914,
Rasht, Iran
\\$^{3}$ Institute for Advanced Studies in
Basic Sciences, P.O.Box 45195-1159, Zanjan, Iran\\
$^{4}$ Max-Planck-Institut f\"{u}r Physik komplexer Systeme,
N\"{o}thnitzer Stra$\beta$e 38, 01187 Dresden, Germany \\
$^{5}$ Department of Physics, Alzahra University,19834, Tehran,
Iran
}
\date{\today}
\begin{abstract}
We study the 1D   ferromagnetic Ising (spin-1/2) model with the Dzyaloshinskii-Moriya (DM)
interaction.  We
analyze the low energy excitation spectrum and the ground
state magnetic phase diagram using the Lanczos method.
The DM interaction-dependency  is calculated for the
low-energy excitation spectrum,  spiral  order parameter and
spin-spin correlation functions. We show that a metamagnetic
quantum phase transition occurs between the ferromagnetic and  spiral  phases. The existence
of the metamagnetic phase transition is confirmed, using the
variational matrix product states approach.
\end{abstract}

\pacs{75.10.Jm Quantized spin models;75.10.Pq Spin chain models }

\maketitle

\section{Introduction}\label{sec1}

Studying one dimensional quantum spin systems have been obtained many interesting results.
The Ising
spin models  pose intriguing theoretical problems because
antiferromagnetic (AF) and ferromagnetic (FM) systems with spin-1/2
have a gap in the excitation spectrum. Therefore they reveal an extremely
rich behavior dominated by quantum effects.
In particular, the spin-1/2
Ising model in a transverse magnetic field (ITF) displays a
pragmatic example of a quantum phase transition.
Theoretically, the ITF problem is exactly solved\cite{Pfeuty70} and
found a phase transition at a finite value of the transverse
magnetic field: $h_{c}$. This is a quantum critical point and
the phase transition is continuous in nature.

The antisymmetric spin exchange
interactions  between
spins, known as the Dzyaloshinskii-Moriya (DM) interaction,
 play an important role in physics of spin
systems\cite{Dzyaloshinskii58,Moriya60}.
The DM interaction idea
originated from the deviation of
experimental data   from the
theoretical predictions, based on the  Heisenberg spin Hamiltonians\cite{Dender97,Sirker04,Sakai94,Chaboussant98,Kageyama99,Cepas01,Jaime04,Zheludev99, Derzhko06}.
Generally the DM interaction between two spins
$S_1, S_2$ can be written as ${\bf D}\cdot ({\bf S}_1 \times {\bf
S}_2)$ with an axial DM vector ${\bf D}$. In actual systems, the
direction of ${\bf D}$ vector is fixed by the microscopic
arrangement of atoms and orbitals. In a spin chain,
the DM vector may vary both in direction and magnitude. However, the
symmetry arguments usually rule out most of possibilities and confine
the theoretical discussion to two principal cases. The first one
is the uniform DM interaction, ${\bf D}=constant$ over the
system\cite{Schotte98} and the second case is the staggered DM
interaction\cite{Oshikawa97} with anti-parallel ${\bf D}$ on
adjacent bonds.
Since the DM interaction is rather difficult to handle
analytically, the effect of this interaction are only partially studied so far.
In this sense we study the Ising  chain (FM) with DM interaction,
that its Hamiltonian (by considering a periodic chain of $N$ sites) is given by
\begin{eqnarray}
{\cal H} = J \sum_{j} S_{j}^{z} S_{j+1}^{z} + \sum_{j}
{\bf D}\cdot ({\bf S}_{j}\times
{\bf S}_{j+1}) , \label{IDF-Hamiltonian}
\end{eqnarray}
where  $\bS_{j}$ is spin-1/2 operator on the j-th site,
and $J>0~(J<0)$ denotes the AF (FM)
coupling constant.
In a very recent work,
using the quantum renormalization group and numerical Lanczos
methods, the ground state phase diagram of the AF
Ising chain ($J>0$) is studied\cite{Jafari08}. It is shown that
the ground state phase diagram consists of AF and
 spiral  phases.

 By considering uniform DM vector as ${\bf D}=D{\hat z}$, and doing
the rotation about $z$ axis
as $S_{j}^{\pm}\rightarrow S_{j}^{\pm} e^{\pm \frac{ij\pi}{2}}$,
the Hamiltonian is transformed to XXZ chain\cite{Alcaraz90,Kaplan83} with an anisotropy
parameter $\frac{J}{D}$, i.e.,
\begin{eqnarray}
{\cal H}^{tr} = D \sum_{j}\left[
(S_{j}^{x}S_{j+1}^{x}+S_{j}^{y}S_{j+1}^{y})+\frac{J}{D}S_{j}^{z}
S_{j+1}^{z}\right].  \label{XXZ-Hamiltonian}
\end{eqnarray}
The
 XXZ chain model was solved by Bethe\cite{Bethe31}, and its  ground state phase diagram
 is well known\cite{Takahashi99}. The Neel regime is
governed by $\frac{J}{D}>1$ and there is a gap in the excitation
spectrum. For $\frac{J}{D} \leq -1$, the ground state is in
the FM phase and there is a gap over the FM
state. In the region $-1<\frac{J}{D}\leq 1$, the ground state of
the system is in the gapless Luttinger  liquid (LL) phase. Thus,
by increasing the $D$ value, the system undergoes a quantum phase transition from the gapped Neel (or
FM) phase to the gapless LL phase at the critical
value $D_c=J$.

In this paper we study an Ising spin-$1/2$ chain with
FM exchange $(J<0)$ and the uniform DM
interaction using the numerical and analytical approaches. In
the forthcoming section,  we apply the Lanczos method to diagonalize numerically
finite chain systems with lengths $N=8, 10,...,24$. Using the
exact diagonalization results, we calculate the spin gap,
magnetization and spin-spin correlations as a function of the DM
interaction. We have also calculated the  spirality  in the ground
state of the system. Based on the exact diagonalization results we obtain the
ground-state magnetic phase diagram of the model. By taking the DM
interaction as the control parameter we show that a metamagnetic
phase transition, can be observed in the  1D ferromagnetic
Ising model. In section III, the observed metamagnetic phase transition from the
numerical lanczos method is confirmed by the analyzing the results
of the variational matrix product states approach.
Finally, we conclude and summarize
our results in section IV.

\section{Numerical results} \label{sec3}

\begin{figure}
\centerline{\psfig{file=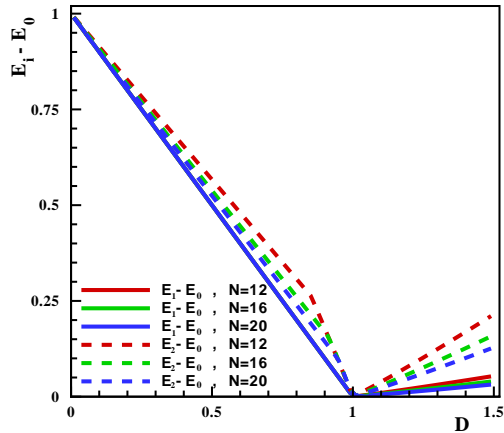,width=3.0in}}
\caption{Difference between the energy of the two lowest
levels and the ground state energy as a function of the DM vector,
for chains with FM exchange ($J=-1.0$) including different lengths $N=12, 16, 20$.}
 \label{energy-gap}
\end{figure}

In this section, to explore the nature of the spectrum and the
quantum phase transition, we use the Lanczos method to diagonalize
 chains with length up to $N=24$.
We have computed the three lowest energy eigenvalues of
chains with FM exchange $J=-1.0$ and different values of the DM vector. To get the energies
of the few lowest eigenstates we consider chains with periodic
boundary conditions.
In Fig.\ref{energy-gap}, we have presented
results of these calculations for the chain sizes $N=12, 16, 20$.
We define the excitation gap as a gap to the first excited state.
It exhibits that
the energy spectrum is gapped at $D=0$, while by increasing the $D$
the energy gap  decreases linearly
and vanishes at $D_c=|J|=1.0$. We got this critical value as an exact value
since there was no finite size correction.
In the region
$D<D_c$,
 the difference between the energy of the first excited state and the
ground state energy shows an universal
linear decrease with increasing DM vector, which is
independent on the chain length (within the used numerical
accuracy). By increasing the DM
vector for $D>D_{c}$, the gap opens again in finite chains, but by
extrapolating  finite size results to $N\rightarrow \infty$, we
found that there is no gap in the spectrum. That is in agreement
with the  transformed Hamiltonian  results (Eq.~\ref{XXZ-Hamiltonian}).
Hence  there are two gapped and gapless phases in the ground state
phase diagram.

\begin{figure}[t]
\centerline{\includegraphics[width=3.0in]{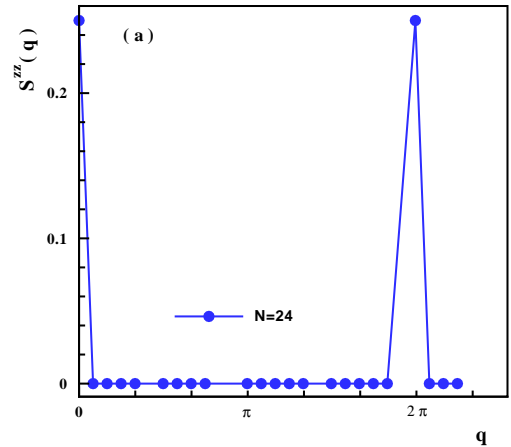}}
\centerline{\includegraphics[width=3.0in]{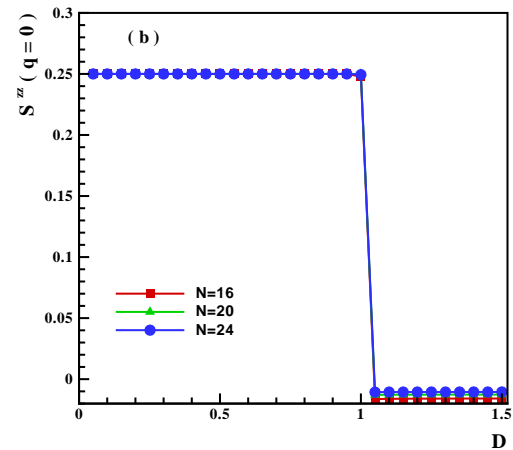}}
\caption{a) The spin structure factor $S^{zz}(q)$ as a function of
 $q$ for FM Ising chain with  DM interaction ($J=-1.0$ and $N=24$).
b) The spin structure factor $S^{zz}(q=0)$ as a function of
the DM interaction $D$ for FM Ising chain ($J=-1.0$) including different chain lengths
$N=16, 20, 24$.  }
 \label{Spin-structure}
\end{figure}

\begin{figure}
\centerline{\psfig{file=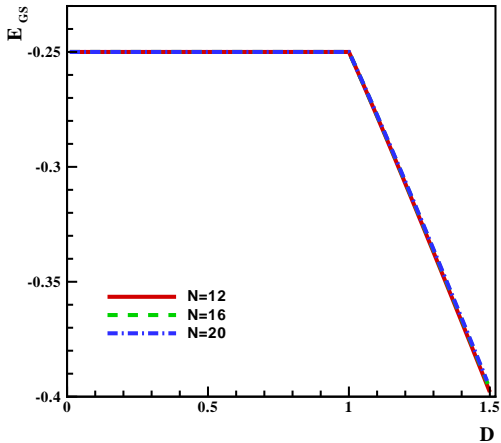,width=3.0in}} \caption{The
numerical results on the ground state energy (normalized by N) of
a FM Ising model with  DM interaction as a function of
$D$ for different chain lengths $N=12, 16, 20$ ($|J|=1.0$). }
\label{EGS}
\end{figure}

To study the magnetic order of the ground state of the system, we
have implemented the Lanczos algorithm on finite chains to
calculate the lowest eigenstate.
The symmetry breaking
cannot occur in finite size systems, thus instead of  the magnetization,
$M^z=\frac{1}{N} \sum_{j} \langle S_{j}^{z} \rangle$, we focus on the spin-spin
correlation functions. The static spin structure factor at
momentum $q$ is defined as
\begin{eqnarray}
S^{z z}(q)=\frac{1}{N} \sum_{n} e^{iqn}\langle
S_{j}^{z}S_{j+n}^{z} \rangle . \label{structure-factore}
\end{eqnarray}
It is known that the spin structure factors give us deeper
insight into the characteristics of the ground state.
The $D$-dependency of the spin structure factor, $S^{zz}(q=0)$,
is qualitatively the same as the uniform
magnetization, $M^{z}$.

\begin{figure}
\centerline{\psfig{file=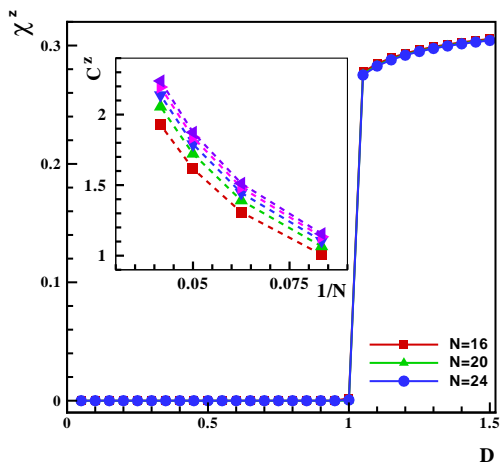,width=3.0in}}
\caption{The spiral order parameter  as a function of the DM
vector for the FM Ising chain ($J=-1.0$). Inset shows the spiral
correlation function, $C^{z}$, as a function of the $1/N$ for different
values of the DM vector: $D=1.1, 1.2, 1.3, 1.4, 1.5$.    }
\label{Chiral}
\end{figure}

In Fig.\ref{Spin-structure}(a) we have plotted the spin structure
factor, $S^{zz}(q)$, as a function of momentum $q$ for a chain
with length $N=24$ and different values of the DM vector in the
region $D<D_c$. We found that in this region the function
$S^{zz}(q)$ is independent of the value of $D$. It shows that
$S^{zz}(q)$ has a very sharp peak at $q=0$, corresponds to the FM
ordering. On the other hand, there is not any peak in the region
$D>D_c$ which confirms that all spins should aline in the $xy$
plane. In Fig.\ref{Spin-structure}(b), we have plotted
$S^{zz}(q=0)$ as a function of $D$ for the chain lengths $N=16,
20, 24$. For $D<D_c$ the spin structure factor, $S^{zz}(q=0)$, is
equal to the value $0.25$, which shows that the ground state of
the system is in the fully polarized FM phase. This quantity is
independent of the chains size, therefore the value of the
magnetization does not change in the thermodynamic limit $N
\rightarrow \infty$, and the FM ordering is a true long range
order  in the region $D<D_c$. One of the most interesting
predictions of this model is that the magnetization as a function
of the DM vector displays a jump for certain parameters. It means
that the spontaneous magnetization, $M^z$, remains at the
saturation value in the region $D<D_c$. But at the critical value
$D=D_c$, the spontaneous magnetization jumps to zero. A rapid
increase (or discontinuity) at critical value of the control
parameter in the magnetization curve is called the metamagnetic
phase transition\cite{Ito92,Kaczmarska95}. This phenomena that
observed in the ground state phase diagram of the 1D frustrated FM
Heisenberg model\cite{Mutter98,Mahdavifar08-1}, has defined the
phase transition  between the FM and AF phases.
 Surprisingly, however, we find that
 the metamagnetic phase transition can be observed between
the FM and  spiral phases in the 1D ferromagnetic
spin-$1/2$ Ising model with DM interaction. In following we draw
a simple physical picture for this phenomena.

The wave function of the ground state in the absence
of the DM interaction has a form
$| \psi_{GS} \rangle=|\uparrow\uparrow\uparrow\uparrow ... \rangle$.
Applying a uniform DM interaction on the FM
state $| \psi_{GS} \rangle$ yields
\begin{eqnarray}
\sum_{j} {\bf D}\cdot ({\bf S}_{j}\times
{\bf S}_{j+1})| \psi_{GS} \rangle=0,
\end{eqnarray}
which shows that in the presence of a DM interaction the ground
state is fully polarized and does not change up to the critical value, $D_c$.
The system is fully FM in $0\leq D <D_c$, and lies
in the subspace $S_{tot}=N/2$ with two-times degeneracy.
But it
becomes an incommensurate state for $D>D_c$, where the energy gap is strongly
suppressed. At the critical point, two distinct
configurations with the energy
$E_{GS}=-\frac{1}{4} N |J|$
are the ground states, where one is fully polarized in $z$
direction.
In
Fig.\ref{EGS}, we have plotted the Lanczos results on the ground
state energy per sites as a function of $D$ for the chain
lengths $N=12, 16, 20$. As we expected in the region of $D<D_c$ the energy
per spin of the ground state  has the constant value $-0.25$ and
independent of the DM interaction ($E_{GS}/N=-|J|/4$).

On the other hand, we have also computed numerically the transverse spin structure factors ($S^{x x}(q), S^{y y}(q)$) for different values of DM vector. Due to symmetry $S^{x x}(q)$ is the same as $S^{y y}(q)$. We found that for the values of the DM vector $D<D_c$, $S^{x x}(q)=S^{y y}(q)=0$ in well agreement with the saturated ferromagnetic phase in the $z$ direction. In the region $D>D_c$, the transverse spin structures showed two peaks at $q=\frac{\pi}{2}, \frac{3 \pi}{2}$, which is a justification of the spiral order\cite{Jafari08}. It is showed that the DM interaction can
induce the  spiral phase in the ground state phase diagram of the
spin systems\cite{Mahdavifar08-2}, which is characterized by the
nonzero value of the  spirality
\begin{eqnarray}
\chi^{z}=\frac{1}{N} \sum_{j}\langle ({\bf S}_{j}\times {\bf
S}_{j+1})^{z}\rangle . \label{spirality}
\end{eqnarray}
One should note that there are two different types of the  spiral
ordered phases, gapped and gapless\cite{Kaburagi99,Hikihara00}.
Therefore, the definition of the spiral
correlation function   as
\begin{eqnarray}
C^{z}=\sum_{n=1}^{N}\langle \chi_{j}^{z}~\chi_{j+n}^{z}\rangle,
 \label{spiral-cor}
\end{eqnarray}
provides further insight into the nature of different
phases.
In Fig.\ref{Chiral}, we have presented results of these
calculations for Ising chains with different lengths $N=16, 20, 24$.
In complete agreement on the results of
magnetization, the  spirality in the FM Ising
chain  shows a plateau in zero value at $D<D_c$.
Which confirms that there is no spiral long range
order in the mentioned regime and the ground state is in the
saturated FM phase. The  spirality  remains close to
zero value in the region $D<D_c$, but at the critical point, it
jumps to a non-zero value. Which is also an indication of the
metamagnetic quantum phase transition that occurs only in the
case of FM Ising chains.
 The inset of the Fig.\ref{Chiral}, shows the spiral
correlation function, $C^{z}$, as a function of the $1/N$ for different
values of the DM vector.
It shows that the spirality  remains non zero in the thermodynamic limit 
for $D>D_{c}$, that corresponds to the spiral long range order in the $xy$ plane.

\section{Variational Matrix Product States approach}

\begin{figure}
\centerline{\psfig{file=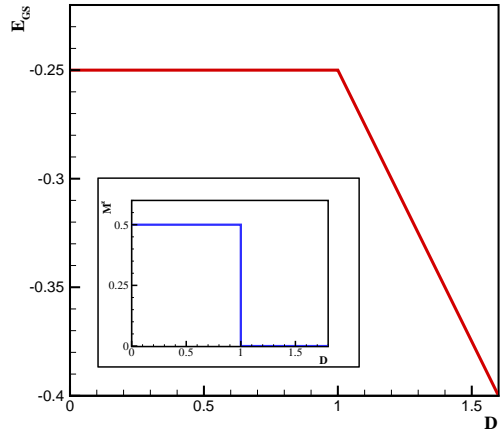,width=3.0in}}
\caption{The ground state energy (normalized by N) of
FM Ising with DM interaction using the variational
matrix product state. Inset shows the spontaneous magnetization
$|M^z|$ as a function of the DM interaction, using the
variational matrix product state ($|J|=1.0$).  } \label{MPS_E0}
\end{figure}
The matrix product state is defined as \cite{Klumper91,Fannes89}
\begin{equation}
   |\Psi\rangle
   =Tr(g_1\cdot g_2\cdots g_{N}),
   \label{E:Psi}
\end{equation}
where the elementary matrix  $g_j$ represents the
matrix-state of the $j$th spin cell. The size of
these elementary matrices depends on the problem. In this paper
the simple one dimensional case
 is considered, i.e., $g_{j}=a_j |\uparrow\rangle_{j}+ b_j|\downarrow\rangle_{j} $. Where $a_j$ and $b_j$ are
probability amplitude   for two configuration of the spin at site
$j$, and $N$ is the  lattice site number. The goal is to determin
the ground state energy of the FM Ising spin system  with DM interaction. In this respect,
the variational energy is obtained by
\bea E_{var}= \langle{\cal H} \rangle = \frac{\langle\Psi|{\cal
H}|\Psi\rangle } {\langle\Psi|\Psi\rangle} =\sum_j
\frac{{\hat{\cal H}_{j,j+1}}}{[ G_{j}\cdot G_{j+1}]}, \eea where
$\langle\Psi|\Psi\rangle = g_{1}^{\dagger}\otimes  g_{1}  \cdots
g_{N}^{\dagger}\otimes  g_{N} = \prod_j G_{j}$,
 and $G_j$ is defined by $ G_{j}=g_{j}^{\dagger}\otimes  g_{j}=|a_j|^2+|b_j|^2$.
Here ${\hat{\cal H}_{j,k}}= J{\hat S^{z}_j}{\hat S}^{z}_k +{\bf D}
\cdot {\bf {\hat S_j}} \times {\bf {\hat S_k}}$, and
 ${\hat S^{\alpha}_j}=g_j^{\dagger}\otimes S_j^{\alpha}g_j$.

The minimum of the variational energy function corresponds to the
ground state energy of the system. Using
 the normalization condition  $\langle\Psi|\Psi\rangle=1$, variational parameters can be mapped to
$a_j=\cos(\theta_j)e^{i\varphi_j}$,
$b_j=\sin(\theta_j)e^{i\varphi'_j}$. Therefore one  can obtain
 ${\hat S^{z}_j}=\cos(2\theta_j)/2$,
${\hat S^{+}_j}=({\hat
S^{-}}_j)^{\ast}=1/2\sin(2\theta_j)e^{i\phi_j}$, where
$\phi_j=(\varphi'_j-\varphi_j)$. By choosing ${\bf D}=D{\hat z}$,
it is easily  found that
\bea E_{var}&=&\frac{1}{4} \sum_j[
J\cos(2\theta_j)\cos(2\theta_{j+1})+
\nonumber\\
&&D\sin(2\theta_j)\sin(2\theta_{j+1})\sin(\phi_j-\phi_{j+1})].
\eea
By minimizing the above equation, the ground state energy
($E_{GS}$) in the FM case $J<0$ shall be achieved. One
can show that the ground state energy has the constant value,
i.e.  $E_{GS}=-N|J|/4$ for $D<|J|$, and  it behaves almost
linearly for  $D>|J|$ as  $E_{GS}=-ND/4$ (please see the
Fig.\ref{MPS_E0}).

The  magnetization can be written  as
$M^z=\sum_j\cos(2\theta_j)/(2N)$, thus by considering the
minimized variational parameters it follows the metamagnetic
phase transition  curve which has shown in the inset of the
Fig.\ref{MPS_E0}, where $|M^z|=0.5$ at $D<|J|$ and zero for the
$D>|J|$. Also the  spirality  is given by
\begin{eqnarray}
\chi^{z}=\frac{1}{4N}\sum_j \sin(2\theta_j)\sin(2\theta_{j+1})\sin(\phi_j-\phi_{j+1}),
\end{eqnarray}
where using above results,  one can obtain $\chi^{z}=0$
for the $D<|J|$, and $|\chi^{z}|=0.25$ for $D>|J|$.

\section{Summary and Conclusion}

In this paper the elementary excitations and the magnetic ground
state phase diagram of  the 1D spin-$1/2$ FM Ising model with the
Dzyaloshinskii-Moriya interaction have been thoroughly
investigated by numerical tools and variational schemes. Using the
analytical and numerical approaches, we have shown that there are
two different phases in the zero-temperature phase diagram of the
model.
To provide a physical picture of the ground state phase diagram
of the model, by a redefinition of the spin variables the model
is mapped onto an XXZ Heisenberg chain. Where the anisotropy
parameter is related to the DM vector, and identified a
commensurate-incommensurate (C-IC) quantum phase transition
between the gapped and gapless phases.
The numerical experiment with high accuracy, has shown that
in the ground state
phase diagram of the FM chain with DM interaction,
there is only one fully polarized FM phase below the
critical value: $D_c=|J|$. However at the critical value, a
metamagnetic phase transition occurs to the  spiral  gapless phase.
Moreover, we used a variational method for evaluating ground
state energy. By considering 1D simple matrix product
state, the zero temperature energy diagram of the ferromagnetic
Ising chain with DM interaction has been obtained. Our data shows
the sudden jumping in the magnetization and  spirality curves at
$D_c=|J|$.

Based on the results we conclude that
properties of the FM and AF Ising
spin-$1/2$ chains with the uniform DM interaction are very different.
In principle, the ground state phase diagram of the FM
chain consists of two phases which are gapped and gapless. In the
case of FM chains, for  $D<D_c$,
there is only one saturated FM phase
and at the critical value, $D_c$, a metamagnetic phase transition
occurs to the  spiral  phase.

\section{acknowledgments}
It is our pleasure to thank T. A. Kaplan  for reading carefully our
manuscript and appreciate his very useful comments. We also  acknowledge  S. D. Mahanti,
and P. M. Duxbury for  useful comments. A. A. would like to thank M. Haque  for valuable comments and useful discussions.


\vspace{0.3cm}

\end{document}